\documentclass[aps,prd,amsmath,showpacs,12pt]{revtex4}
\usepackage{epsfig}
\usepackage{graphicx}
\usepackage{amsmath}
\usepackage{amssymb}
\usepackage{caption2}
\usepackage{rotating}
\usepackage{subfigure}
\usepackage{bm}
\usepackage{color}

\usepackage[colorlinks = true,
linkcolor = magenta,
urlcolor  = blue,
citecolor = red,
anchorcolor = blue]{hyperref}
\newcommand{\beq}{ \begin{equation}}
\newcommand{\eeq}{\end{equation}}

\begin{document}

\title{The Quantum  Statistical Approach to Parton Distributions
\vskip 0.1cm 
updated with recent experimental data}

\email{claude.bourrely012@orange.fr}

\author{Claude Bourrely}

\affiliation{Centre de Physique Th\'eorique, CNRS, Marseille,
Universit\'e de Toulon, France}

\begin{abstract}
The Quantum Statistical Parton Model (QSPM) has been successful 
over the years to  explain a 
important set of unpolarized and 
polarized experimental data 
on proton parton distribution functions (PDFs).
With the advent of the Marathon and SeaQuest experiments, an 
updated version of the quark PDFs obtained in the
framework of QSPM is presented. 
Moreover, in order to clarify the role of the thermodynamical potentials,
the main parameters of the model,
we examine the dependence of the nucleon
entropy on these potentials. 
We find that the values of
these potentials lead to a maximal value for the nucleon entropy.
\end{abstract}


\maketitle

\section{Introduction}
\setcounter{page}{1}
The (QSPM) has a long history which begins
with an unpolarized version \cite{BAL}.

Since the  origin of the Quantum statistical parton ind functions
including polarized effects are based on the Pauli principle 
which implies a Fermi-Dirac statistics
for the fermions and a Bose-Einstein one for the gluon
were widely used to explain a great number of results obtained
in unpolarized and polarized experiments.
The first version \cite{BBS} published in the year 2002
contains already the basic elements of a simplified formulation,
for instance, the simultaneous treatment of the evolution of
parton helicity distributions, 
which is differeht of the calculations
made in classical  (PDFs) models.
In the version \cite{BBS} the simplified expression of the strange quark and the polarazed 
gluon are not adapted to describe certain new polarized processes.

Nevertheless, we have obtained the striking result, the ratio 
$\bar d / \bar u$ increases at large Bjoken $x$  and Q$^2$ = 54GeV$^2$
in contradiction with the result of the E866 NuSea experiment \cite{e866} which 
shows a decrease, one can notice that at that time all the parton models 
were in agreement with these data.
This property of the (QSPM) to show an increase comes from the
constraint due to  a simultaneous fit of unpolarized and polarized data.
To conclude, we waited 17 years and the  results of the E906 
experiment \cite{e906} to confirm our original result.

After this first version of the (2002) model was released plenty of new experiments
have been performed leading us to build a modified version of the (QSPM)
with a more general (PDFs) expressions in order to describe these great set 
of unpolarized and polarized data \cite{BS15}. We succeeded in the year 2015
to achieve this project
with a second version 'BS15' of the (QSPM)
published in the year 2015.
To give an idea of the results, 2042 unpolarized and polarized experimental 
data points were fitted at NLO with a good accuracy of a $\chi^2/pt = 1.276$.

Later on, new measurements have been obtained by Marathon, SeaQuest, and
STAR collaborations, \cite{marathon}\cite{STAR1}\cite{Do1},
so in order to maintain the quality of the (QSPM) new PDFs are defined
in sec. 2. A fit of experimental data is made in sec. 3 which gives the values of
the PDFs parameters.
In sec. 4 some comparison ot statistical model predictions with experimental 
data are shown. 
In particular, 
the Isospin Asymmetry in the proton sea
measured by  the SeaQuest experiment in sec. 5 and
the proton  Spin Asymmetry in sec. 6.
 Sec. 7 presents the calculation of
the proton entropy and a search for an optimum associated
with the thermodynamical potentials is performed.

\section{Definitions of the quantum statistical parton distributions}
The fermion distributions for the quarks  $q$ and $\bar q$
 are given by the sum of two terms,
a quasi Fermi-Dirac function also called the non diffractive part 
which is helicity dependent
and a helicity independent diffractive contribution:
\begin{equation}
xq^h(x,Q^2_0)=
\frac{A_{q}X^h_{0q}x^{b_q}}{\exp [(x-X^h_{0q})/\bar{x}]+1}+
\frac{\tilde{A}_{q}x^{\tilde{b}_{q}}}{\exp(x/\bar{x})+1}~,
\label{equa1}
\end{equation}
\begin{equation}
x\bar{q}^h(x,Q^2_0)=
\frac{{\bar A_{q}}(X^{-h}_{0q})^{-1}x^{\bar{b}_ q}}{\exp
[(x+X^{-h}_{0q})/\bar{x}]+1}+
\frac{\tilde{A}_{q}x^{\tilde{b}_{q}}}{\exp(x/\bar{x})+1}~,
\label{equa2}
\end{equation}
defined at the input energy scale $Q_0^2=1 \mbox{GeV}^2$. We note that the diffractive
term is absent in the polarized quark helicity distributions 
$\Delta q = q^{h=+} - q^{h=-}$ and also in the quark
valence contribution $q - \bar q$.

In Eq.~(\ref{equa1}) the multiplicative factors 
$X^{h}_{0q}$ 
in the numerator and in Eq.~(\ref{equa2})
$(X^{-h}_{0q})^{-1}$ 
in the denominators of the non-diffractive parts of the $q$'s and $\bar{q}$'s
come in fact from the extension of the statistical model to
the tranverse momentum dependence TMD-PDFs \cite{tmd}.
The parameter $\bar{x}$ plays the role of a {\it universal temperature}.\\
The parameters
 $X^{\pm}_{0q}$ are the two {\it thermodynamical potentials} of the quarks
$q$, with helicity $h=\pm$. They represent the fundamental characteristics of
the (QSPM)  that we call now the 'B26' version to sepate it  from the 
 BS15 version.
Although we consider light quarks for the moment,
these formulas are valid for the 6 flavors.
The difference with the  previous (QSPM) version the BS15 relies on the parameters
$A_{q}$ and  $b_q$ which are now different for the $u,~d$ quarks 
introducing two more parameters.
 
One would like to stress that  a  fit of the statistical model
 determines the parameters of
the non diffractive unpolarized component $q_{ND} = q^+ + q^-$.
The evolution at larger $Q^2$ will produce different unpolarized
and polarized PDFs.
The HOPPET evolution program  \cite{hoppet} allows us to compute simultaneously
in the fitting process 
the unpolarized and polarized parton distributions which imposes
a strong constraint on the parameters.
Let us remark that the unpolarized quark contains in addition
a diffractive contribution absent in the polarized ones. 

In summay,
for a given flavor $q$ the corresponding quark and antiquark distributions
involve the free parameters, $X^{\pm}_{0q}$, $A_q$, $\bar {A}_q$,
$\tilde {A}_q$, $b_q$, $\bar {b}_q$ and $\tilde {b}_q$, 
whose active number is reduced when the valence sum rule is imposed, 
$\int (q(x) - \bar{q}(x))dx = N_q$, where $N_q = 2, 1, 0 ~~\mbox{for}~~ u, d, s$, respectively.

The other free parameters are
$\tilde {A}_u = \tilde {A}_d$, $b_u = b_d$, $\bar {b}_u = \bar {b}_d$
and $\tilde {b}_u = \tilde {b}_d$. 

For the strange quark and antiquark
distributions, the simple choice is the one made  in Ref.  \cite{BS15}.
For the gluons one considers the black-body inspired expression
\begin{equation}
xG(x,Q^2_0) = \frac{A_Gx^{b_G}}{\exp(x/\bar{x})-1}~,
\label{eq3}
\end{equation}
a quasi Bose-Einstein function, with $b_G$ previbeing the only free parameter,
since $A_G$ is determined by the momentum sum rule.
In our previous work \cite{BBS} it was assuming that, at the input
energy scale, the polarized gluon distribution vanishes, so it was set to
\begin{equation}
x\Delta G(x,Q^2_0)=0~.
\label{eq4}
\end{equation}
However, as a result of the present analysis of a much larger set of very
accurate unpolarized and polarized DIS data, one must give up this simple form.
A new expression is assumed:
\begin{equation}
 x\Delta G(x,Q^2_0) = \frac{\tilde A_G x^{\tilde b_G}}{(1+ c_G
x^{d_G})}\!\cdot\!\frac{1}{\exp(x/\bar x - 1) } \,.
\label{polglue}
 \end{equation}
It is clear that we don't have a serious justification of the functional form
of Eq. \ref{polglue}. The above expression shows that it is strongly
related to  $G(x,Q^2_0)$ and therefore constructed by means of a Bose-Einstein
distribution with zero potential. 
Actually since 
$\Delta G(x,Q^2_0)=P(x)G(x,Q^2_0)$ 
a simpler expression would be $P(x) = Ax^b$, but an additional
term  in the denominator is needed in order to get a reasonable fit
of the polarized DIS data.
Also to insure that positivity is satisfied we must have
$|P(x)| \leq 1$, which is automatically fullfilled by construction.
A posteriori an other justification of the polarized gluon expression is given 
by the calculation of the jet asymetry $A_{LL}^{jet}$ \cite{bs15}.
An equivalent situation is observed in the calculation of this asymmetry
with polyninominal polarized gluon, see Vogelsang {\it et al.} \cite{vogelsang}.

\section{The statistical model fit of experimental data}
The experimental data used in the fit are listed in tables 1 and 2 of Ref. \cite{BS15}
 \cite{ccfr01}-\cite{cms14} completed with the ratio results $F2N/F2P$ of the 
Marathon experiment \cite{marathon} 
and the ratio  $\sigma(pd)/\sigma(pn)$ measured by SeaQuest  \cite{Do1,Do}, 
notice that the other functions given by SeaQuest are deduced from this 
cross sections ratio,
they  will be discussed later.
\begin{table}[htb]
\begin{center}
\caption{Summary of the fit results}
\label{table1}
\begin{tabular}{|c|c|c|}
\hline 
 &  $\chi^2$ & $\chi^2/pt$\\
\hline
B26 (2206 pts) & 3208 & 1.45 \\
F2N/F2P (22 pts) & 20.1 & 0.96  \\
$\sigma(pd)/\sigma(pn)$ (6 pts) & 10 & 1.66 \\
\hline
\end{tabular}
\end{center}
\end{table}
These results characterize an excellent fit of the global data 
including Marathon and Sea-Quest.

\begin{figure}  
\begin{center}
\includegraphics[
height=7.0cm,width=6.0cm]{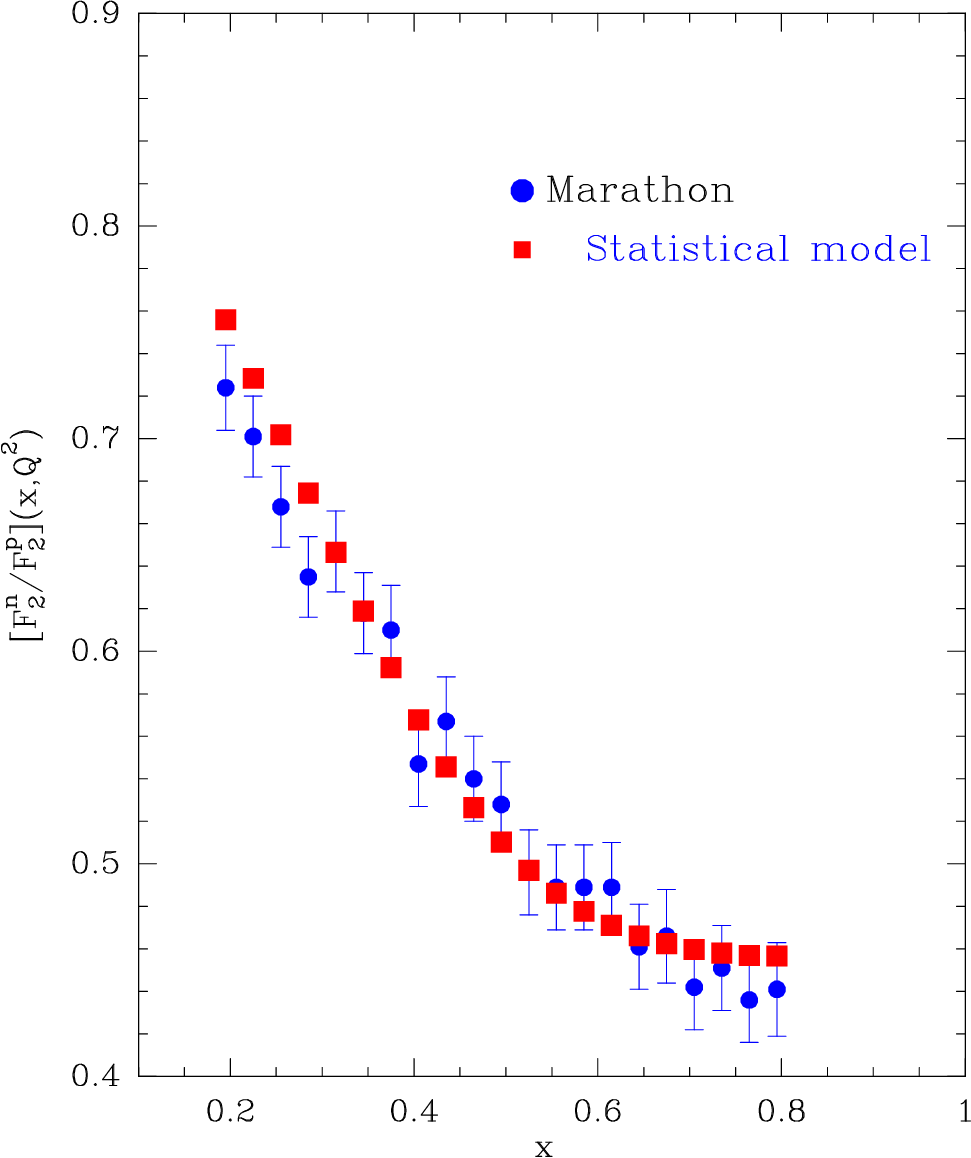}
\caption{\baselineskip 1pt\ref{fig11}
 Plot of the  ratio $F2n(x,Q^2)/F2p(x,Q^2)$ for each couple $(x,Q^2)$ 
 by the MARATHON data \cite{marathon}, blue circles,the red squares are the fitted values.}
\label{fig1}
\end{center}
\end{figure}

\begin{figure}   
\begin{center}
\includegraphics[height=6.0cm,width=6.5cm]{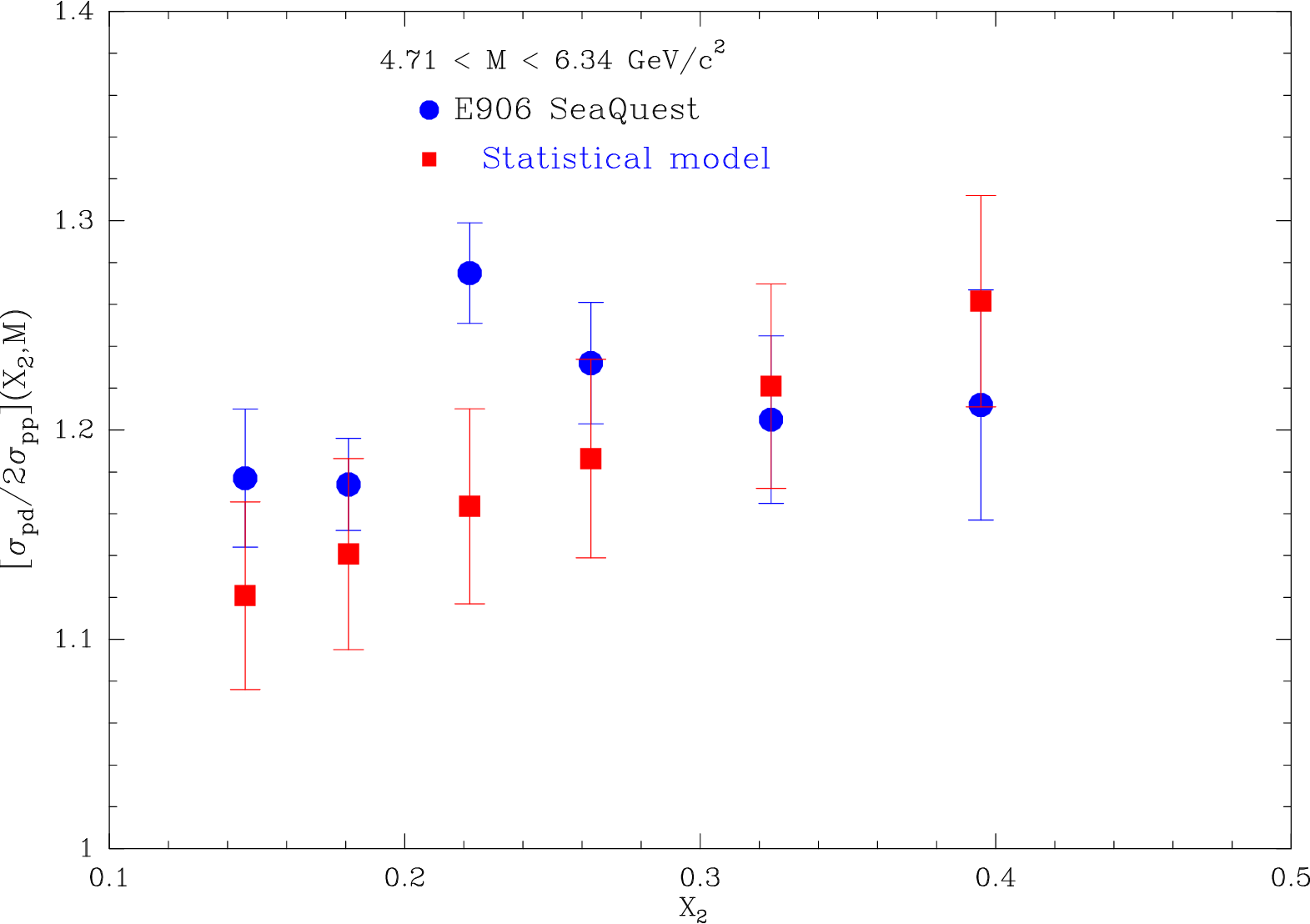}
\caption{\baselineskip 1pt
 The fitted cross section ratio $\sigma_{ p d} / 2\sigma_{p p}$.}
\label{fig2}
\end{center}
\end{figure}

The list of the fitted parameters values reads:
\begin{eqnarray}
\nonumber
A_u = 2.2749, ~\bar {A}_u = 0.18045,~b = 0.5025,\\
A_d = 2.6272, ~\bar {A}_d = 0.4053,\\
~\bar {b}_q = 0.7944,~\tilde A = 0.1490, ~\tilde b= 0.0328,
~\bar x = 0.10315.
\label{eq9}
\end{eqnarray}
and four $u$, $d$ potentials
\begin{eqnarray}
\nonumber
X_{0u}^+= 0.43378,~X_{0u}^-= 0.31409,\\ X_{0d}^+= 0.17173,
~ X_{0d}^-= 0.3008.
\label{eq10}
\end{eqnarray}
The other parapeters read:
\begin{eqnarray}
\nonumber
A_s = 26.6192,~b_s = 0.14115,~\bar {A}_s = 0.02373, \\ 
\bar {b}_s = 3.9503,~\tilde A_s = 10.2769, ~\tilde b_s= 19.2161,\\
\label{eq11}
\end{eqnarray}
\begin{equation}
X_{0s}^+= 0.00462,~X_{0s}^-= 0.011898.
\label{eq12}
\end{equation}
Finally, in the gluon sector, the following parameters are obtained:
\begin{eqnarray}
\nonumber
A_G = 26.6192,~b_G = 0.9900,~\tilde {A}_G = 164.7831\\
\tilde {b}_G = 0.06268,~c_G = 0.006, ~d_G = 5.4209.
\label{eq8}
\end{eqnarray}
 The fitted ratio $F2n(x,Q^2)/F2p(x,Q^2)$ measured by Marathon \cite{marathon}
and the SeaQuest fitted ratio $\sigma_{ p d} / 2\sigma_{p p}$
are plotted in Figs. \ref{fig1}, \ref{fig2}.

\section{Predictions of experimental data with the
fitted  statistical model}
\begin{figure} 
\begin{center}
\includegraphics[height=6.0cm,width=6.5cm]{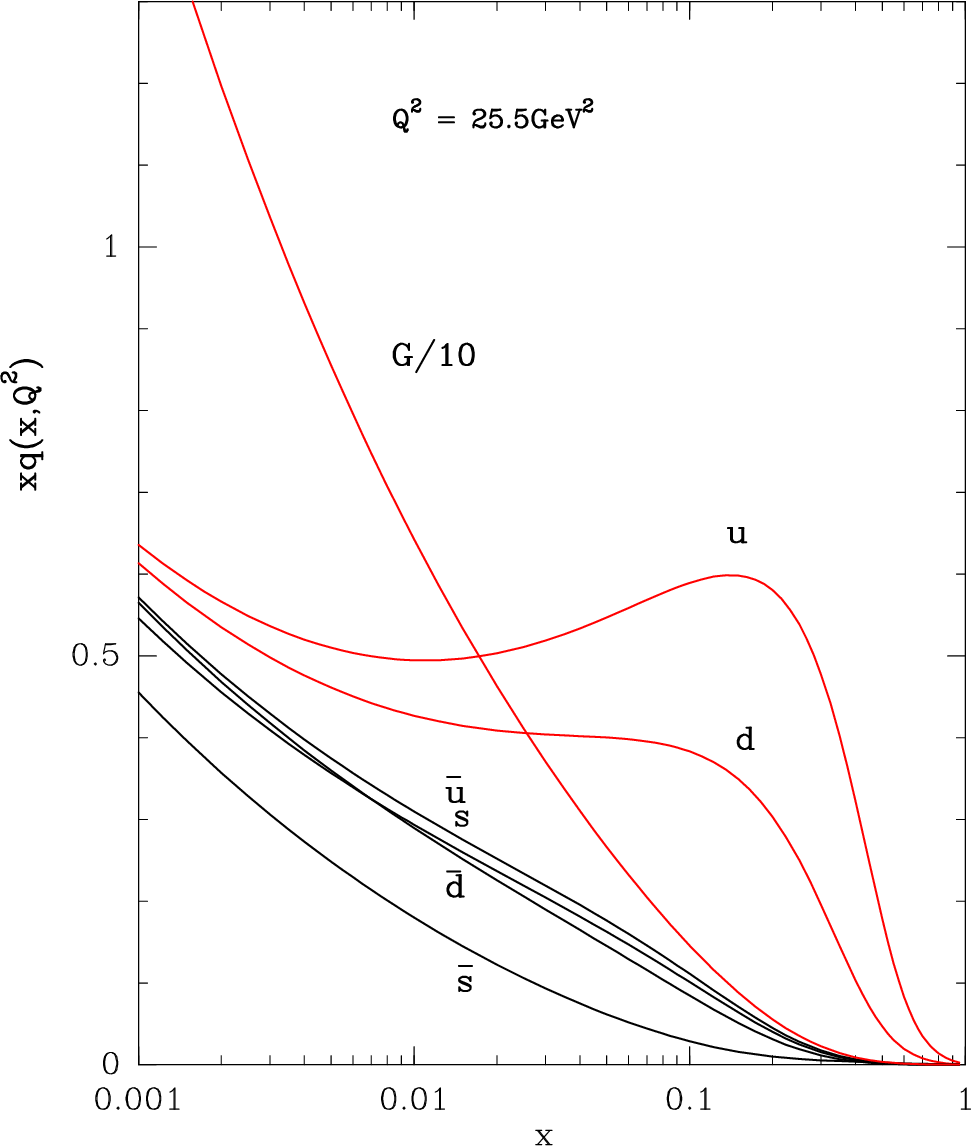}
\caption{\baselineskip 1pt
 Plot of the unpolarized PDFs at $Q^2 = 25.5\mbox{GeV}^2$.}
\label{fig3}
\end{center}
\end{figure}

\begin{figure} 
\begin{center}
\includegraphics[width=6.0cm]{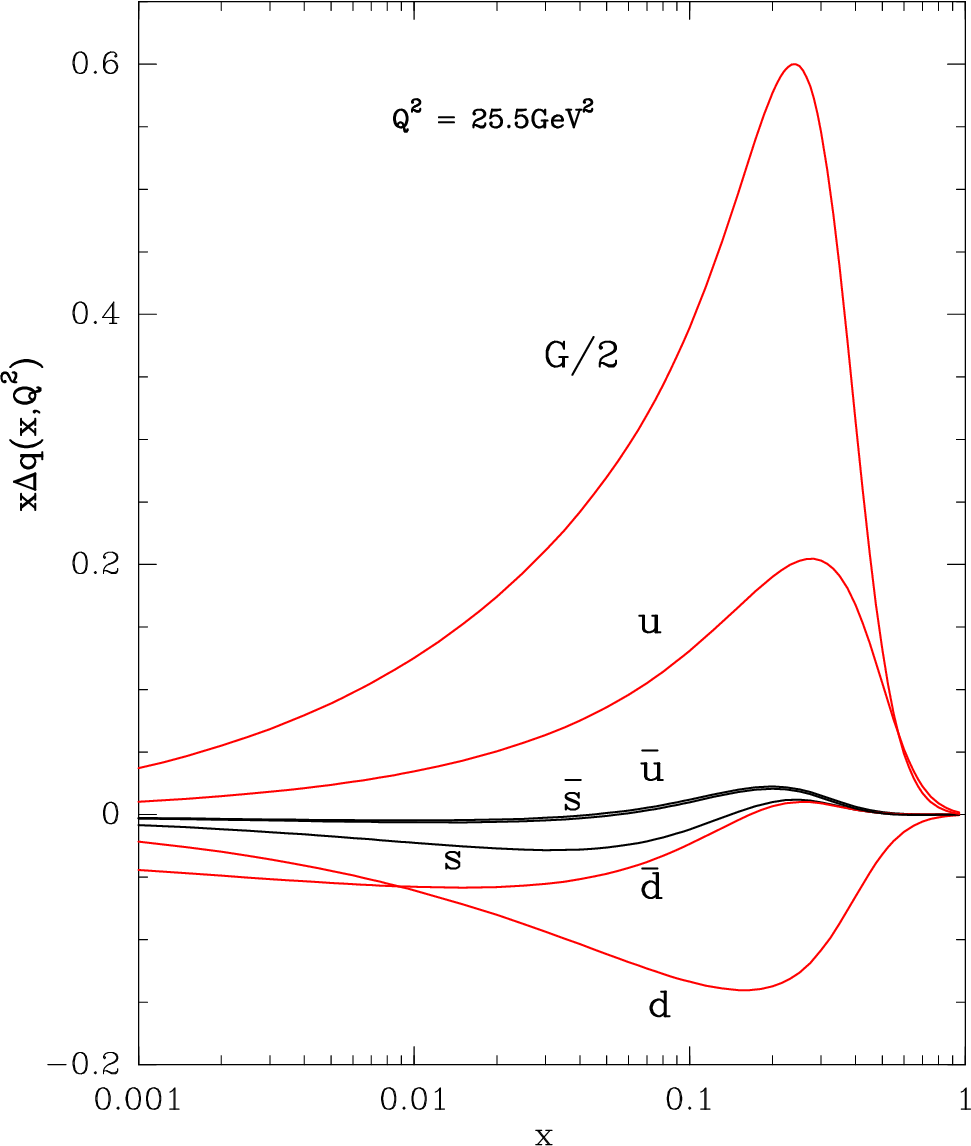}
\caption{\baselineskip 1pt
 Plot of the polarized PDFs at $Q^2 = 25.5\mbox{GeV}^2$.}
\label{fig4}
\end{center}
\end{figure}

\begin{figure}  
\begin{center}
\includegraphics[width=6.0cm]{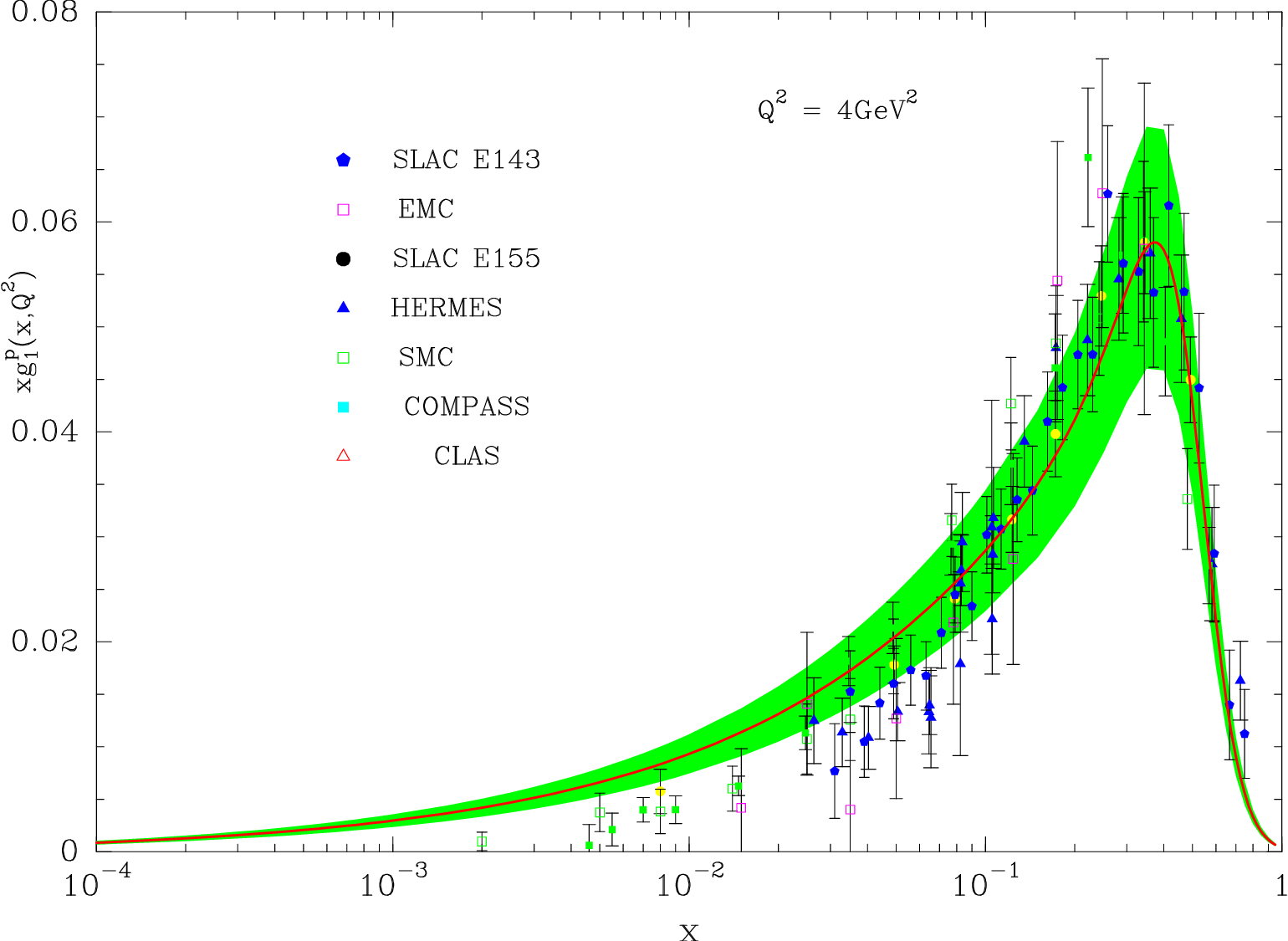}
\caption{\baselineskip 1pt
 Plot of the polarized structure function $g_1^p(x,Q^2)$ at $Q^2 = 4\mbox{GeV}^2$.}
\vskip -0.5cm
\label{fig5}
\end{center}
\end{figure}

In this section some significant figures  illustrate the
properties of the statistical model. 
Parton distributions are plotted in Figs. \ref{fig3}, \ref{fig4}.
In the polarized case two different structure funtions 
$g_1{p,n}(x, Q^2)$ are shown in  Figs. \ref{fig5},\ ref{fig6} making in evidence the simultaneous
possibility to handle both unpolarized and polarized processes.
\begin{figure}  
\begin{center}
\includegraphics[width=6.0cm]{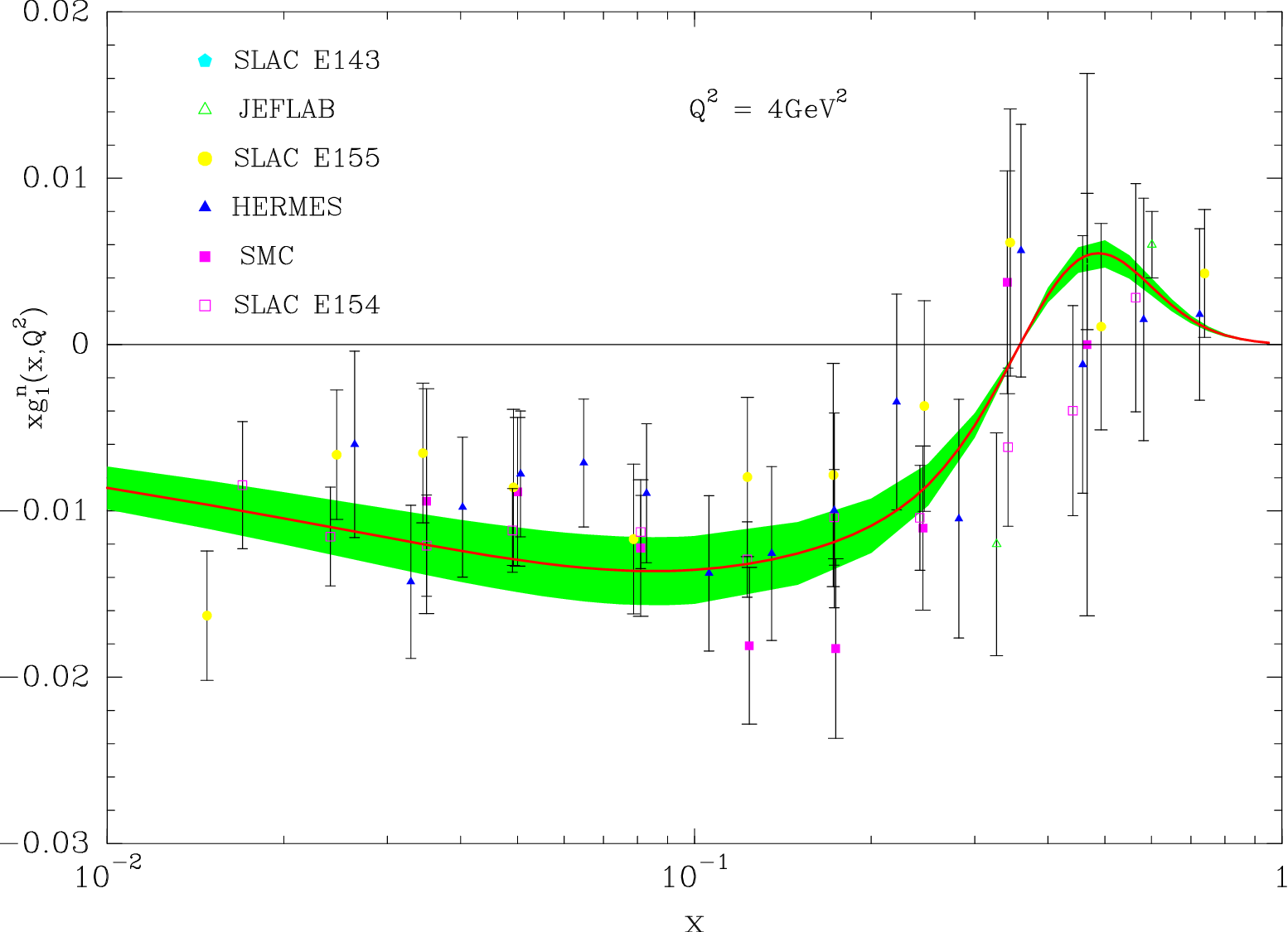}
\caption{\baselineskip 1pt
 Plot of the polarized structure function $g_1^n(x,Q^2)$ at the same $Q^2$.}
\label{fig6}
\end{center}
\end{figure}

\section{The isospin asymmetry in the proton sea
measured by  the SeaQuest experiment}
The SeaQuest experiment \cite{STAR1,STAR2} has measured the isospin 
asymmetry.
They also give an estimate of the moments (see Table VIII of ref. \cite{Do})
\begin{equation}
\int_{0.13}^{0.45}  [\bar d(x) -\bar u(x)] dx = 0.0159
\label{mn01}
\end{equation}
the calculation of this moment with (QSPM) gives 0.0128
and
\begin{equation}
\int_{0.13}^{0.45} x  [\bar d(x) -\bar u(x)] dx = 0.00318
\label{mnt1}
\end{equation}
the calculation of this moment with (QSPM) gives 0.00267, 
these two  results are perfectly compatible
with the experimental data taking into account the experimental errors.
In the following figures \ref{fig7}, \ref{fig8}, 
our excellent results are  
compared with their data:
\begin{figure}
\begin{center}
\includegraphics[width=6.0cm]
{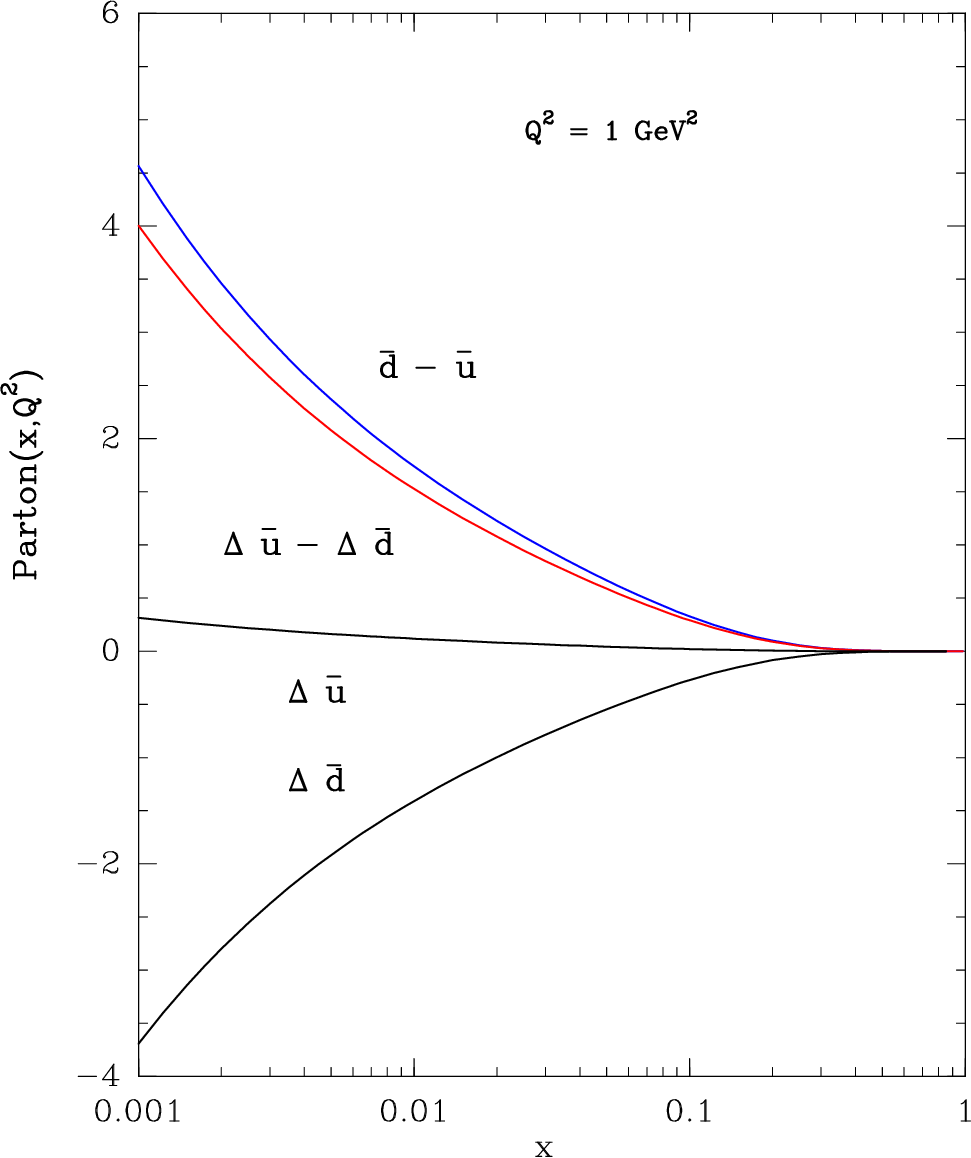}
\caption{\baselineskip 1pt
 Plot of the inequalities between polarized and unpolarized partons distributions,
at $Q^2 = 1 \mbox{GeV}^2$.
$\Delta \bar d < 0 < \Delta \bar u < \Delta \bar u - \Delta \bar d < \bar d - \bar u  $.}
\label{fig7}
\end{center}
\end{figure}
\begin{figure}
\begin{center}
\includegraphics[width=6.0cm]
{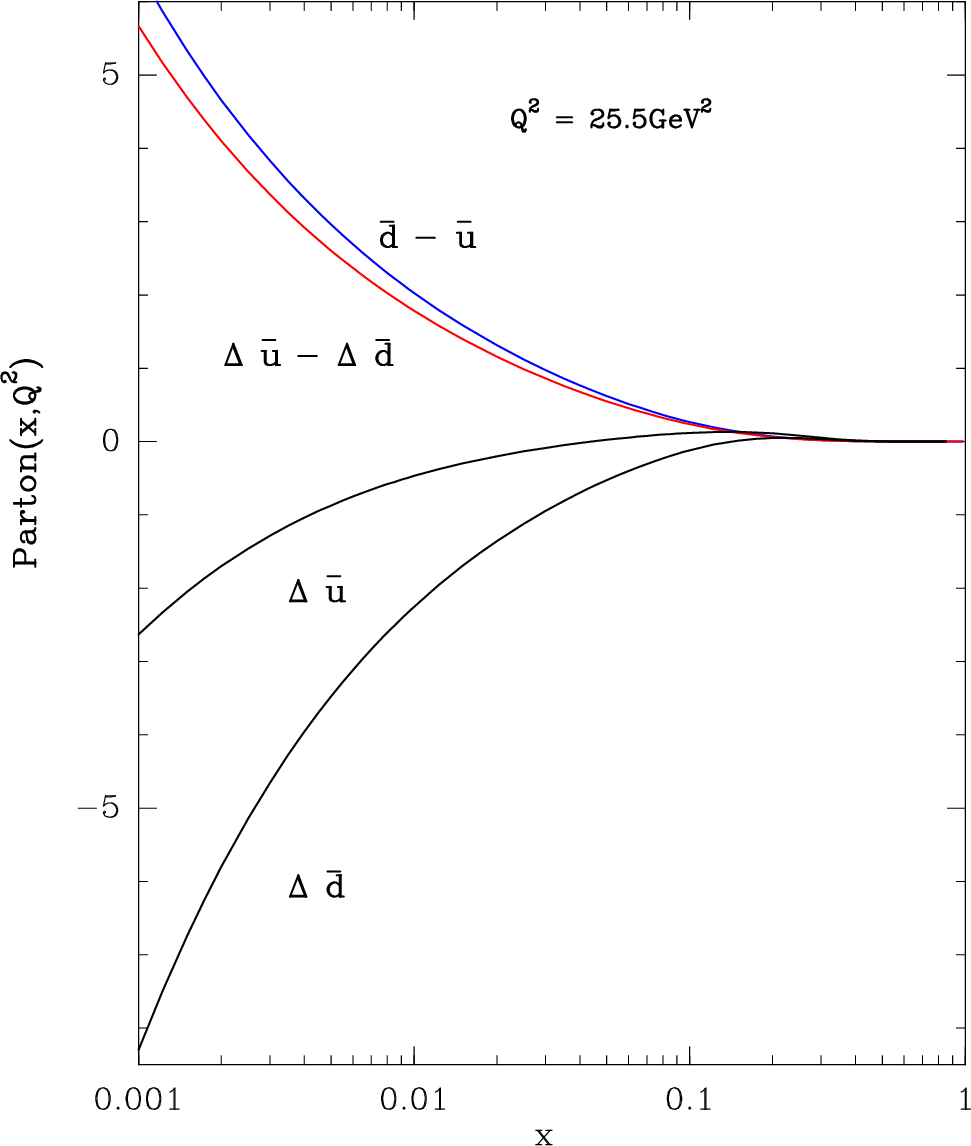}
\caption{\baselineskip 1pt
 The same inequalities as in figure 7 at $Q^2 = 25.5 \mbox{GeV}^2$.}
\label{fig8}
\end{center}
\end{figure}
Some other results linking inequalities between polarized and unpolarized 
distributions are shown in Figs. \ref{fig9}, \ref{fig10}.
\begin{figure}
\begin{center}
\includegraphics[width=6.0cm]{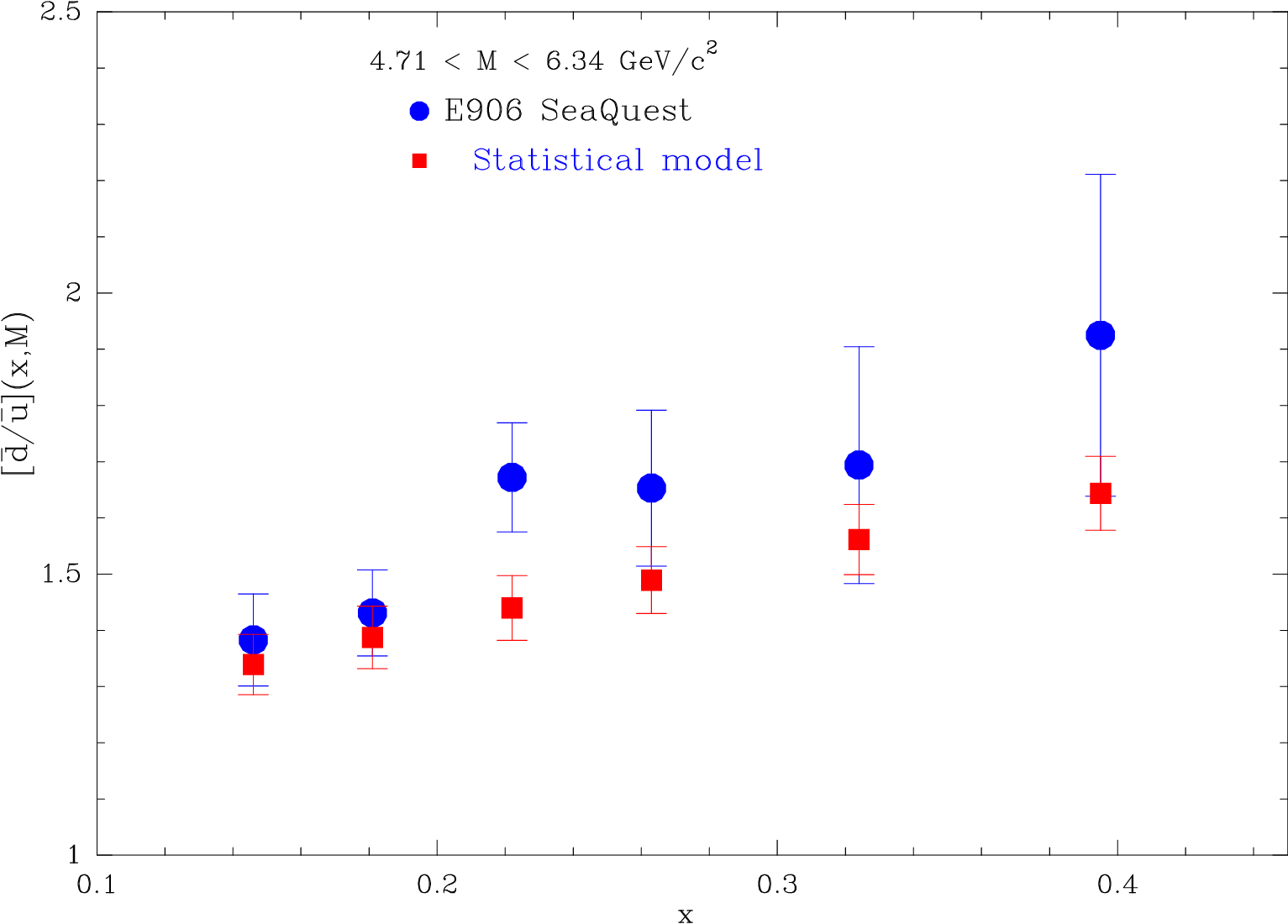}
\caption{\baselineskip 1pt
 $[\bar d / \bar u]$ ,
blue circles SeaQuest data. red squares the statistical model.}
\label{fig9}
\end{center}
\end{figure} 
\begin{figure} 
\begin{center}
\includegraphics[width=6.0cm]{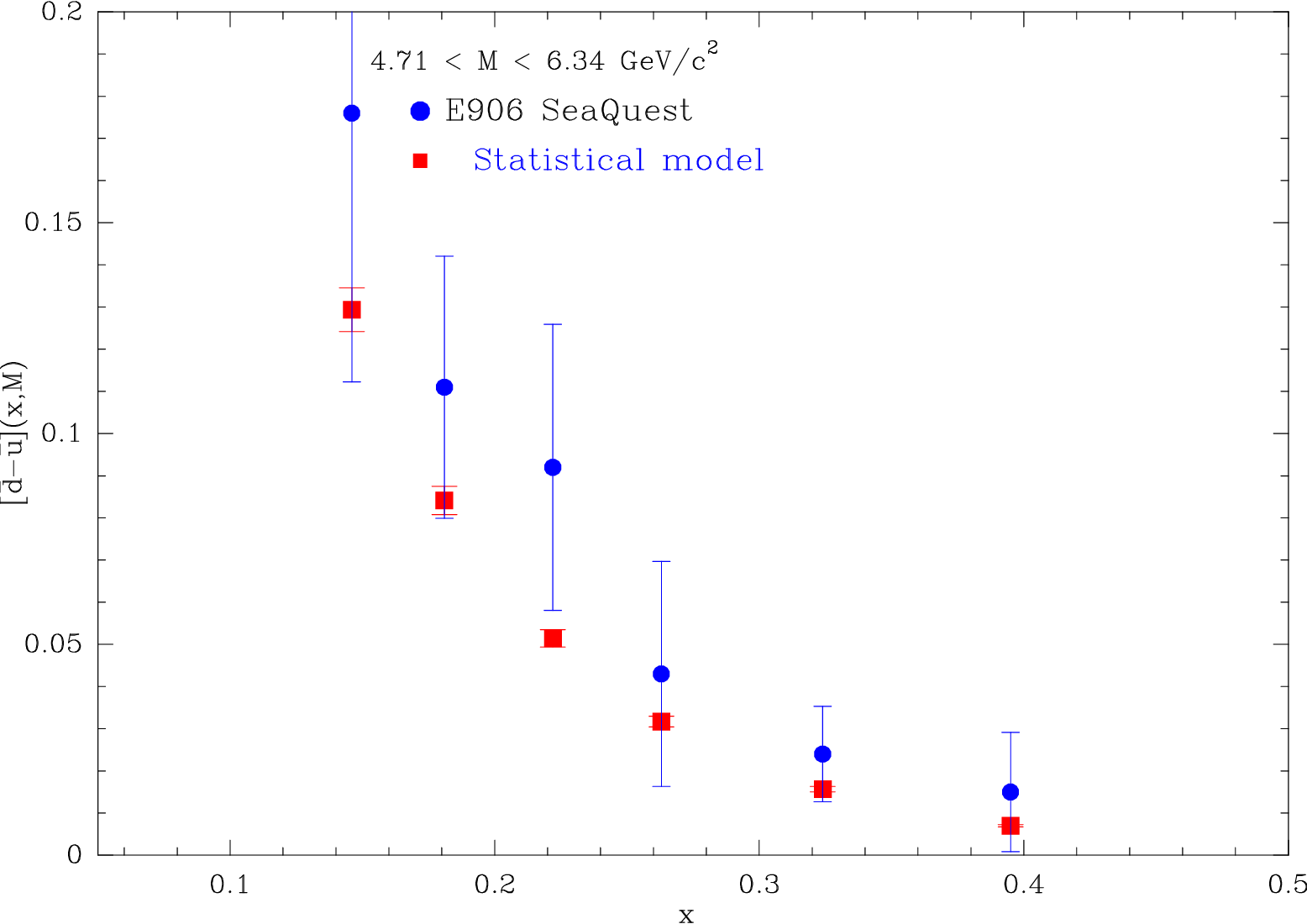}
\caption{\baselineskip 1pt
 $[\bar d - \bar u]$ , blue circles SeaQuest data. red squares the statistical model.}
\label{fig10}
\end{center}
\end{figure} 

\section{The spin asymmetry  in a proton measured by STAR experiment}
\begin{figure} 
\begin{center}
\includegraphics[width=6.0cm]{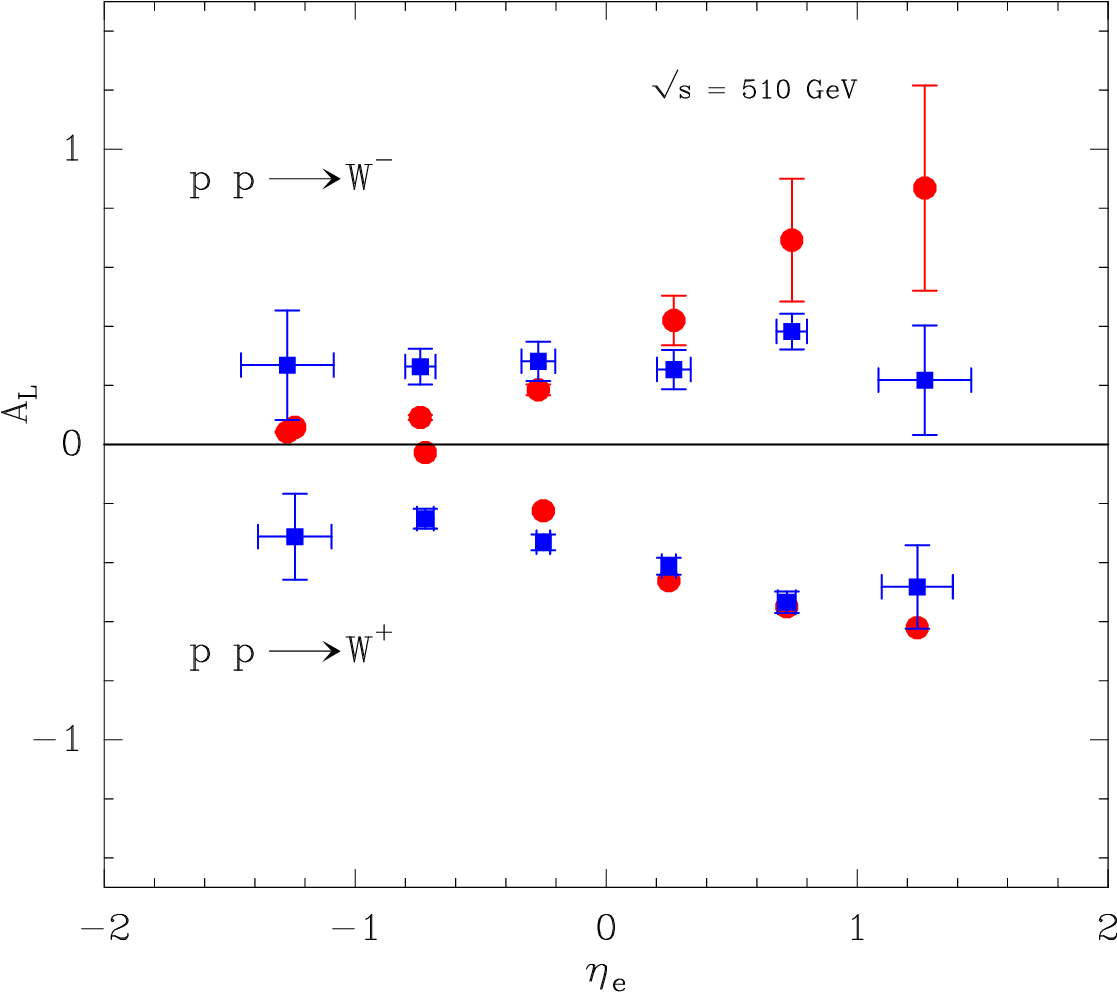}
\caption{\baselineskip 1pt
The measured  helicity asymmetries $A_L$ for
charged-lepton production at BNL-RHIC from STAR \cite{STAR1}, through
production and decay of $W^{\pm}$-bosons versus $\eta_e$  the charged-lepton
rapidity.}
\label{fig11}
\end{center}
\end{figure}

Finally we consider the process $\overrightarrow p p\to W^{\pm} + X  \to
e^{\pm}+X$, where the arrow denotes a longitudinally polarized proton and the 
outgoing
$e^{\pm}$ have been produced by the leptonic decay of the $W^{\pm}$-boson. The
parity-violating  asymmetry is defined as
\begin{equation}
A_L = \frac{d\sigma_+ - d\sigma_-}{d\sigma_+  + d\sigma_-}~.
\label{AL}
\end{equation}
Here $\sigma_h$ denotes the cross section where the initial proton has
helicity $h$. 
$A_L$ was measured recently at RHIC-BNL \cite{STAR1} and the results are
shown in Fig. \ref{fig11}.
The $W^-$
asymmetry is very sensitiveref to the sign and magnitude of $\Delta \bar u$, so
this is another successful result of the statistical approach.
\clearpage

\section{The proton entropy and the  nature of a produced optimum}
An attempt to explore a new property  of the unpolarized PDFs is made
by the calculation of  a physical quantity  like the entropy.
The entropy can be calculated according to the definition given in
Ref. \cite{cleym12} Eq. (16)
\begin{eqnarray}
E(Q^2,x) = -\sum_i \left[x q_i(Q^2,x) \ln{(xq_i(Q^2,x))} \newline
+(1- xq_i(Q^2,x)
\ln{(1- xq_i(Q^2,x))}\right]\,
\label{entropy}
\end{eqnarray}
where the sum runs over the quark components.
One first remark that the vanishing
of $xq_i(Q^2,x)$ for $x = 1$,  implies  that $E(Q^2,x) = 0$ in this limit.
A proposal to compute the entropy for the states generated by
the quarks constituants of the proton, the neutron and the antiproton 
wich correspond to the states  $|2u +d>$, $|u +d +s>$ and $|2\bar u +\bar d>$,
 at a fixed $Q^2 = 10\mbox{GeV}^2$ as a function of $x$ is shown
in Figure 12.
The first state is largely dominant over the last ones
which seems to reflect the importance of matter over anti-matter.
The curves shown in  Figure 12  have been calculated with
the values of the potentials obtained from a fit of experimental data discussed 
above, then a question arises, 
what is the origin of these values, does exist a possibility to obtain them
independently of experimental data? In the PDFs formulas described above we have
introduced the following  parameters: a normalization $A$, 
a power $b$,
a temperature $\bar x$ and the potentials. 
For a matter of simplification 
in the calculation let us assume 
that the following parameters $A,~b,~\bar x$ are held fixed to their fitted
values and consider now only  the thermodynamical potentials as free parameters
which will be determined from a calculation of an optimal  entropy
 (\ref{entropy}) 
for a given value of $x$ and $Q^2$.
In complete generality all the parameters of the model should have been 
considered as free, but due the complexity of the computation we restrict our
search only for the six potentials defined as the master parameters of the model.
For this purpose one considers $E(Q^2,x)$ given by Eq. (\ref{entropy}) as an 
{\it objective function} which depends on $u$, $d$, $s$ quarks subjects to the
constraints due to sum rules
\begin{eqnarray}
&& 0 < X_{0q}^h < 1,\quad\quad \int u_v(x) dx = 2, 
\quad\quad \int d_v(x) dx = 1, 
 \\
&& \Delta u(x) \ge 0, \quad\quad \Delta d(x) \le 0,  \quad\quad \Delta s(x) \le 0,
\\
&& \int (xu(x) +xd(x)+ xs(x))dx \le 1, 
\quad\quad \int (s(x) -\bar s(x))dx =0 \,.
\label{constrain}
\end{eqnarray}
The goal is to solve the sytem of equations 
(\ref{entropy})-(\ref{constrain}) with respect to the thermodynamical
potentials (supposed to be unknown) associated with $u$, $d$ and $s$.
The optimization is performed with the NLOPT software \cite{nlopt}, which involves
an objective function, some constraints and their gradients with respect
to the parameters.
In addition, to confirm the results a brute-force method is also applied, it
consists to find the maximum of the entropy by varying the parameters 
in a range defined as $\pm 50\%$ of the fitted parameters values. 
At the fixed $Q^2 = 10\mbox{GeV}^2$
a set of 20 $x$ values in the range
$10^{-3} < x < 1$ are selected, the solutions  for the optimal entropy are
shown in  Fig. 12 as circles for the state $|2u+d>$,
and squares for the state $|u+d+s>$, one observes that their values
are close to the solids curves. 
These results show that the parameters obtained by this method have the
same values (with an error around 2\%)
as the original ones obtained from a data fit, so the entropy obtained from
experimental data satisfies an optimal principle.
\begin{figure}
\begin{center}
\includegraphics[width =6.0cm]{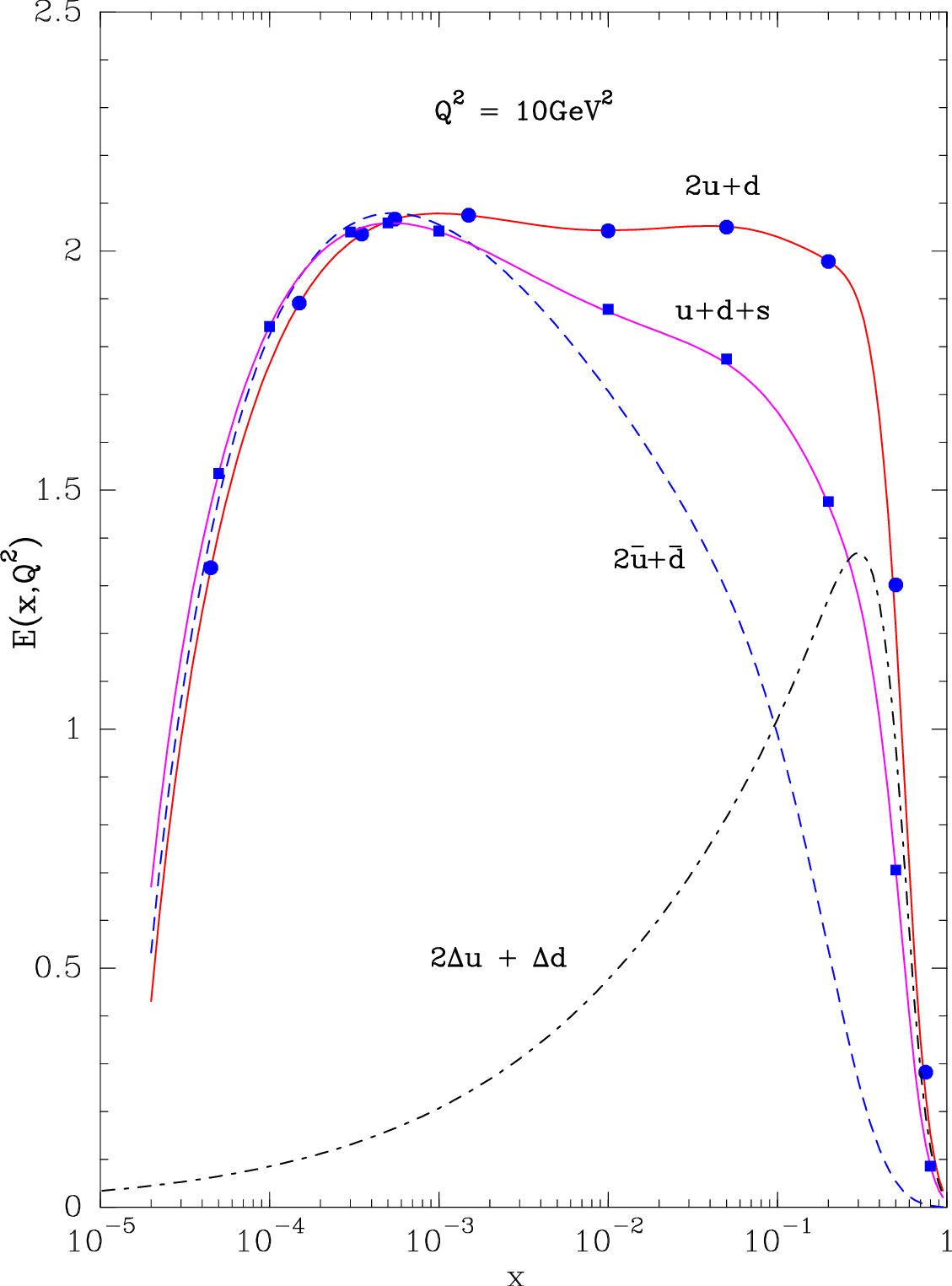}
\caption{\baselineskip 1pt
Entropy at $Q^2 = 10\mbox{GeV}^2$
as a function of $x$ for the states $|2u +d>$, $|u +d +s>$,
$|2\bar u +\bar d>$ and $|2\Delta u + \Delta d>$, 
calculated with the fitted parameters.
The optimal solutions of the entropy correspond to circles for $|2u+d>$,
and squares for $|u +d +s>$.}  
\label{fig12}
\end{center}
\end{figure}
Also  in Fig. \ref{fig12} the produced entropy for polarized state is drawn
 as an illustration with a dash-dotted curve, the values are much smaller than in
the case of an unpolarized proton as expected.
As a conclusion
the calculation of the entropy obtained in an independent way from experiment has
a consequence that the quarks PDFs obtained from a fit correspond to
a maximum entropy value,  the structure functions must share 
in some way the same property. 
Using the same method
as above, by considering again the thermodynamical potentials as free parameters
in a certain range of values a search for a maximum of the structure functions like 
$F_2^p$ and $g_1^p$ at a fixed $x$ and $Q^2$
 could be envisaged. 
In the above results the quark distributions are the essential source
of information to obtain an optimum such a property should be reflected 
in the quarks themselves. 
 Practically, in order to produce a true optimization one needs first 
a multi-dimensional optimization software, 
second a very powerfull computer 
to handle this kind of calculation.
\footnote{The required tools for these calculations were not available to the author.}

\section{Conclusion}
The version of the Quantum Statistical model
presented here is deduced from the 2015 approac
by a slight modification of the normalization factor of $u$, $d$ quarks.
The fit of experimental data made with the present  statistical parton model
defined at an input scale $Q^2$ = 1 $\mbox{GeV}^2$ has
produced values of the structure functions which are
in good agreement with data and in a limited sense are optimal functions with
respect to the values of some main parameters, precisely the 
 thermodynamical potentials.
In some sense one can expect that ``any new'' experimental unpolarized
or polarized measured structure functions data could also be optimal
as they agree with the prediction of the statistical model values.\\ 
All the results  described in this approach give a
strong support to the (QSPM) in this improved formulation,
 so there is no need to work with an approximated PDFs expressions
just for one specific process.
In the present aproach  only light quaks are considered 
but the PDFs expressions 
so far defined can be easily extended to the heavy quarks, 
$c,~b,~t$ which are involved in LHC experiments. 
Also do not forget that the knowledge of unpolarized helicity PDFs
allows us to deduce in a very efficient approximated way
the polarized ones and so it is able to predict the values 
of polarized structure functions.

\section{Acknowledgement}
I thank Prof. Jen-Chieh Peng for interesting discussions on the SeaQuest results.
I am redevable to Prof. Daniel Gandolfo of the Toulon Universit\'e for efficient
help on compuing facilities.\\
In memory of Jacques Soffer, co-author of the BS15 version, who passed away in 2019.\\
%

\end{document}